\documentstyle[epsfig]{elsart}

\def \OM       {{$\Omega_M$}}
\def \snap     {{\it SNAP\ }}
\def \snapns   {{\it SNAP}}
\def \hst      {{\it HST\ }}
\def \hstns    {{\it HST}}
\def \jdem     {{\it JDEM\ }}
\def \jdemns   {{\it JDEM}}
\def \wmap     {{\it WMAP\ }}

\def\arcsec    {\hbox{$^{\prime\prime}$}}

\def\aj        {{AJ}}

\def\apj       {{ApJ}}
\def\apjl      {{ApJ}}
\def\apjs      {{ApJS}}

\def\aap       {{A\&A}}

\def\mnras     {{MNRAS}}

\def\procspie  {{Proc.~SPIE}}

\begin{document}
\begin{frontmatter}
\title{ Exploring Dark Energy with SNAP}
\author{G. Aldering\thanksref{SNSF}}
\address{Physics Division, Lawrence Berkeley National Lab, Berkeley, CA}
\thanks[SNSF]{The work described here is supported by the DOE Office of
Science and by NASA.  The author thanks his collaborators in \snap and
the SCP for their valuable contributions to this work.}
\begin{abstract}
The accelerating expansion of the Universe is one of the most
surprising and potentially profound discoveries of modern cosmology.
Measuring the acceleration well enough to meaningfully constrain
interesting physical models requires improvements an order of
magnitude beyond on-going and near-term experiments. The
Supernova/Acceleration Probe has been conceived as a powerful yet
simple experiment to use Type~Ia supernovae and weak gravitational
lensing to reach this level of accuracy. As fundamentally different
causes for the acceleration map into very small differences in
observational parameters for all relevant cosmological methods, control
of systematics is especially important and so has been built into the
\snap mission design from the very beginning.

Though focused on the study of the accelerating Universe, the overall \snap
instrument suite is quite general and able to make unique contributions
to a wide variety of astronomical studies. The baseline satellite
consists of a 2-m anastigmat telescope, with a 0.7 square degree focal
plane paved with optical and NIR imaging arrays.  Spectroscopy can be
obtained using a high-throughput low-resolution optical+NIR integral field
spectrograph. The baseline science programs will result in a 15 square
degree ``deep field'' having temporal coverage every 4 days 
and summing to $m_{AB}\sim30.3$ in all colors --- to
be used for discovery and follow-up of some 2000 Type~Ia supernova in the
range $0.1<z<1.7$ --- and a wide area survey spanning 1000 square degrees
and reaching $m_{AB}\sim27.7$ in all colors --- to be used to measure the
weak lensing power spectrum well into the non-linear regime. A panoramic
survey covering 10000 square degrees to $m_{AB}\sim26.7$ in all colors is
also possible. This baseline dataset represents a gold mine for archival
astronomical research and follow-up with JWST, while guest observer
survey programs will substantially broaden the impact that \snap will have.
\end{abstract}
\end{frontmatter}
\section{Introduction} 

\snapns\footnote{snap.lbl.gov; see also \cite{omni}} 
is a space telescope mission concept aimed
at measuring the time variation in the equation of state of the dark
energy responsible for the accelerating expansion of the universe.
\snap initially focused on the use of Type~Ia supernovae as a means of
measuring the expansion history of the universe through the
luminosity-distance relation. However, the \snap instrument suite
proved to be quite generic, allowing, for example, the inclusion of
gravitational weak lensing as another means of constraining the
properties of dark energy. Indeed, \snapns's very powerful wide-field
imager and the interest in using it for complementary science has in
many ways motivated and shaped this ``Wide-Field Imaging from Space''
conference.

\section{Cosmology with Supernova from the Ground and Space} 

Before looking at the exciting discoveries that \snap holds for the
future, it is worthwhile to consider recent past and on-going work to
constrain the nature of dark energy.  In the six years since the
discovery of the accelerating expansion of the universe a wealth of new
observations have confirmed and further refined the initial discovery.
In concert with newer data from Type~Ia supernovae (including
constraints on flatness from the CMB and on \OM\ from large-scale
structure and clusters), the statistical significance of the
acceleration is now $>10\sigma$. Just as importantly, the equation of
state of the dark energy is now constrained to be $w<-0.7$ (assuming the
equation of state is constant).

Space-based observations have played an increasing role in cosmological
measurements of Type~Ia supernovae. This is simply due to the much
smaller sky noise present in a point-source photometry aperture in
space compared with the ground --- especially at redder optical
wavelengths.  The discovery papers \cite{Perl99,Riess98} contained just
four supernovae measured with \hstns. Samples totaling 39 Type~Ia
supernovae, segregated into early, middle and late host-galaxy
morphologies from an \hst Snapshot survey, were shown to support an
accelerating expansion \cite{Sullivan03}.  Precise corrections for dust
extinction of eleven more supernovae made possible by accurate \hst
photometry were shown to support the accelerating expansion and
provided tighter constraints on the dark energy equation of state
\cite{Knop03}. A sample of high-redshift supernovae reaching $z\sim1.6$
has shown that the supernovae follow the transition from deceleration
to acceleration expected as the universe goes from matter domination to
dark energy domination \cite{Riess04,Knop03,Tonry03}.  These
space-based measurements have significantly bolstered both the case for
dark energy and the efficacy of Type~Ia supernovae as distance
indicators.  Accordingly, today the most precise and highest-redshift
supernova measurements --- several dozen so far --- are most often
obtained with \hstns.

\begin{figure}
\epsfig{file=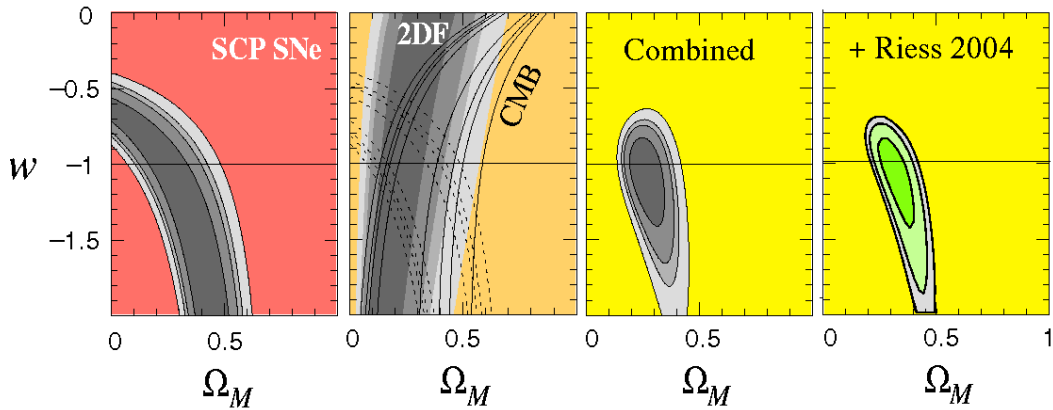,angle=0.0,width=\linewidth}
\caption{The most recent constraints on the dark energy equation of
state, $w$.  A constant equation of state and flat topology are
assumed. Inner and outer contours delimit 68\% and 99\% confidence,
respectively.  On the left, the constraints from the Supernova
Cosmology Project sample \cite{Perl97,Perl98,Perl99,Knop03} are shown.
These include eleven supernovae with very well-measured colors from
\hstns.  The next panel shows independent constraints from \wmap
\cite{Spergel03} and from the 2dF redshift survey \cite{Percival02}.
The third panel shows the results of combining these independent
measurements. The right-most panel shows the effect of the improved
constraints on \OM\ from the Type~Ia supernovae in the deceleration
epoch resulting from the addition of supernovae from the Higher-$Z$
Supernova Search Team \cite{Riess04}.}
\end{figure}

An example of the state of the art in space-based supernova cosmology
is a joint endeavor involving the Supernova Cosmology Project
(SCP{\footnote{supernova.lbl.gov}}) and the Higher-$Z$ Supernova Search
Team (HZSST) which around the time of this conference was conducting a
search spanning four epochs, each covering fifteen pointings covering
the northern GOODS field. The searches are separated by 45 days ---
roughly equal to the time from explosion to maximum light in the
restframe of $z>1$ Type~Ia supernovae.  The search uses the Advanced
Camera for Surveys (ACS) with the 850~nm long-pass filter in order to
reach redward of the restframe UV cut-off exhibited by Type~Ia
supernovae. A short exposure in the F775W 
filter is added in order to reject some fraction of
Type~II supernovae, which in the first 10 days have much more UV flux
than Type~Ia supernovae.  The depth of the GOODS field at optical
(\hstns) and infrared (ground-based) wavelengths means that photometric
redshifts are known for many supernova host galaxies. In addition, the
galaxy colors at the location of the supernovae provide a handle on
whether or not star-formation is likely to be ongoing, which can help
eliminate likely core-collapse supernovae.  All of this advance
information is used to help select one or two likely Type~Ia supernovae
suitable for follow-up.  Each team has three target-of-opportunity
triggers and 60 orbits for near-infrared follow-up observations with
NICMOS and slitless spectroscopy with ACS.  For these searches
each supernova lightcurve and its associated search costs roughly 30
\hst orbits.

At the time of this conference the SCP had netted a probable $z\sim1.7$
Type~Ia supernova from the first of the four search epochs. The
discovery image is show in Figure~2. Since then, several more supernova
have been found as part of this program.  Unfortunately, all of these
supernovae have just the bare minimum of data needed to place them on
the Hubble diagram, leaving little extra information for testing for
systematic effects in the measurement. A satellite such as \snap will
increase the number of such supernovae by orders of magnitude, but will
also have much higher data quality and systematics controls.

\begin{figure}
\epsfig{file=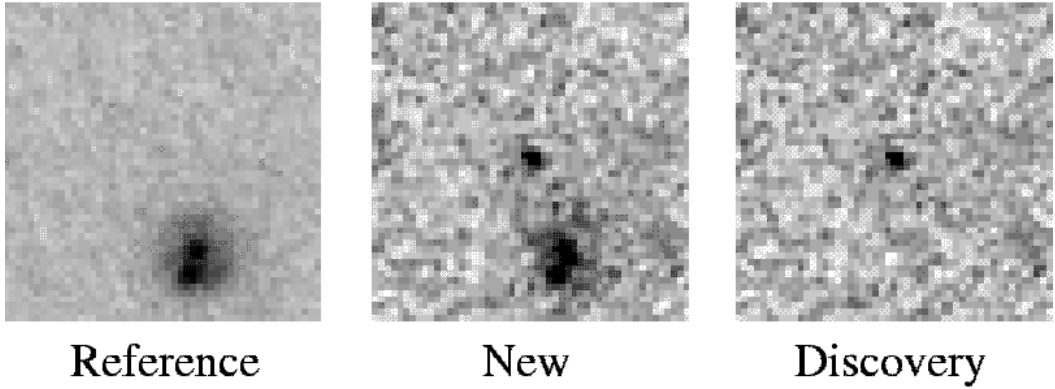,angle=0.0,width=\linewidth}
\caption{Discovery image of a candidate $z\sim1.7$ Type~Ia supernova
from a search of the HDF-N with \hst by the SCP. The image on the left
is the reference image constructed from the GOODS program. The central
panel shows the discovery image, obtained in one \hst orbit in the
F850LP filter. The image on the right is the difference of the
discovery and reference images, isolating the new supernova. The
redshift is based on photometric redshifts kindly provided by GOODS.
\hst F775W images and follow-up ACS slitless spectra do not show blue
light that would be expected if the supernova were of Type~II.}
\end{figure}

\section{Enabling Better Type~Ia Standardization} 

While Type~Ia supernovae have shown tremendous value in revealing the
presence of dark energy and exploring its nature, we should be
concerned that this purely empirical technique will reach a systematics
floor. The single-parameter light-curve width correction method (plus
extinction correction) reduces the scatter in Type~Ia supernovae at
peak to somewhere in the range 0.10--0.15 mag
\cite{Tripp98,Phillips99}.  However, there is nothing special about the
physical conditions in a supernova when it appears its brightest.
Indeed a newly developed technique, CMAGIC, standardizes Type~Ia
supernovae using the color and magnitude roughly 15 days after maximum,
when the supernova reaches a color of $B$-$V$=0.6 \cite{Wang03}. This
method has been shown to reduce the scatter to below 0.10~mag. Thus, it
is possible that Type~Ia supernovae are more homogeneous at a given
color rather than at a given lightcurve epoch.  Many more observations
of nearby supernovae, focusing on accurate colors with well-determined
uncertainties will be a great boon in improving the standardization of
Type~Ia supernovae. By lowering the remaining scatter, stronger limits
are also set on any possible remaining systematics in the Type~Ia
methodology.

\begin{figure}
\epsfig{file=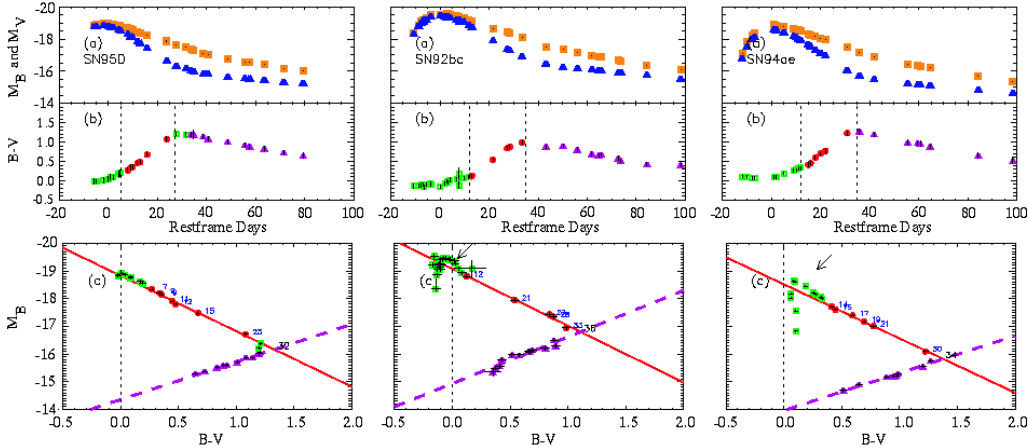,angle=0.0,width=\linewidth}
\caption{Illustration of the CMAGIC Type~Ia supernova standardization
technique for three well-observed Type~Ia SNe. The upper row shows the
$B$ and $V$ lightcurves, the middle row shows the $B$-$V$ color curves,
and the bottom row shows $B$ versus $B$-$V$ color.  Classical
techniques (stretch, $\Delta$m$_{15}$, MLCS) use the lightcurves and
color curves versus time, whereas CMAGIC uses brightness versus color,
which has an amazingly linear region starting about a week after
maximum light.  (There is also a linear region at late times, during
the nebular phase.) The brightness when $B$-$V$=0.6 is an excellent
standard candle \cite{Wang03}.}
\end{figure}

\section{From Science Objectives to Project Design} 

The objective of the \snap mission is to provide the best possible
constraint on the time-evolution of the dark energy equation of state.
The only demonstrated experimental techniques are astronomical, and of
these, the luminosity distance from Type~Ia supernovae is by far the
most mature.  Therefore, it was natural to design a mission focused on
supernovae, but as we shall see, the resulting design is sufficiently
general to perform exceptionally well for gravitational weak lensing.

Pushing the Type~Ia supernova method to its limit requires thousands of
supernovae spanning a wide range in redshift. The present \snap design
targets 2000 normal Type~Ia supernovae from redshift 0.1 to 1.7.  It is
extremely critical that the photometry be of high quality continuously
from low to high redshift, not only to realize the statistical power of
the sample but to avoid introduction of systematics errors
\cite{Kim04}.  Due to the increasing brightness of the terrestrial
night sky at redder wavelengths where higher redshift Type~Ia
supernovae must be observed, coupled with increasing atmospheric water
absorption at redder wavelengths, the necessary data quality can be
obtained only from space.  The broad redshift coverage needed to
control systematics further requires detectors at both optical and
infrared wavelengths.  Telescope optical design and spacecraft size and
mass constraints point to a dedicated 2-m telescope\footnote{For
diffraction-limited photometry of faint point sources, exposure time
goes as the 4th power of the telescope aperture} with a field of view
several hundred times larger than that possessed by \hstns, populated
with visible and near-infrared detectors.  Further, in order to
determine the supernova type, and possible sub-class, a sensitive
low-resolution spectrograph is required.  Systematics control enabled
by spectrophotometric spectra, as well as allowance for spacecraft
pointing margin, call for an integral field unit spectrograph.

\begin{figure}
\epsfig{file=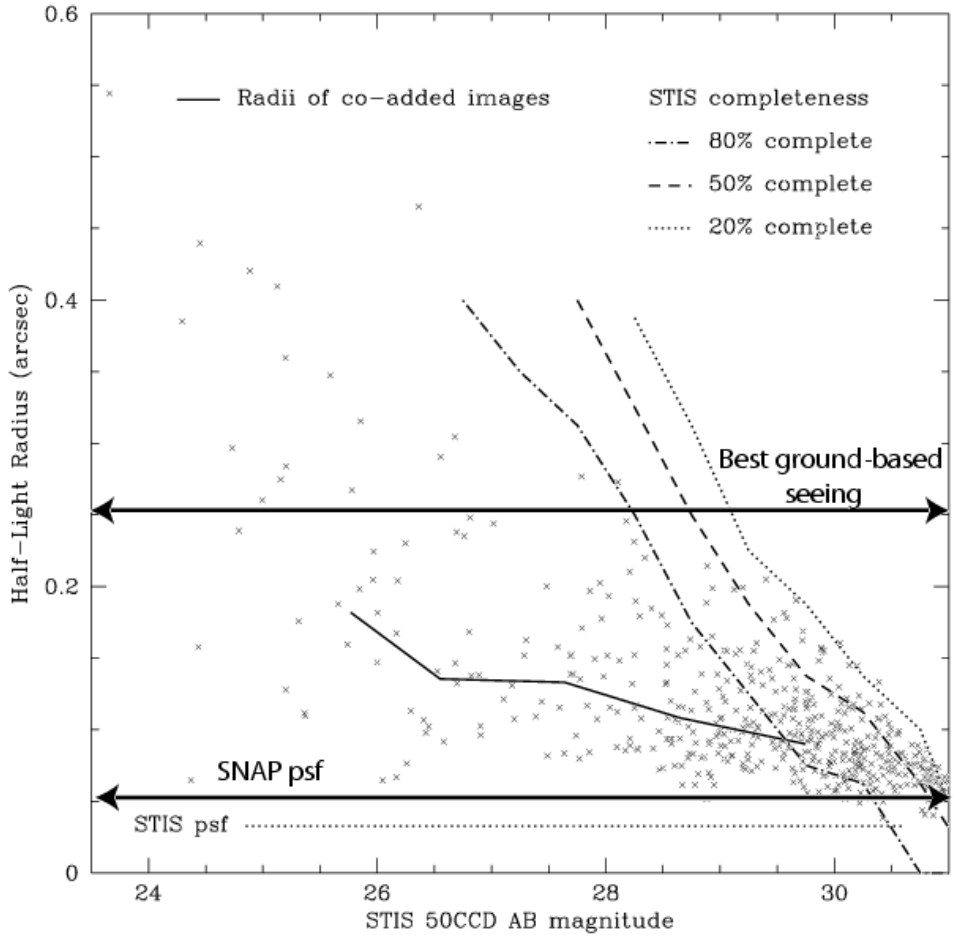,angle=0.0,width=\linewidth}
\caption{Measurements of the sizes of galaxies versus magnitude
\cite{Gardner00}, illustrating the large reservoir of faint, distant
galaxies which can be spatially resolved from space but not from the
ground.}
\end{figure}

\section{The Power of Weak Lensing from Space} 

Gravitational weak lensing measurements have become increasingly robust
over the past decade. Long viewed as a means of constraining \OM, more
detailed theoretical studies --- several developed within the \snap
collaboration --- indicate that precision weak lensing measurements
have the capability to place constraints on the dark energy equation of
state similar to those possible with Type~Ia supernovae. In particular,
weak lensing cross-correlation cosmography --- measuring the
association between foreground template mass distributions and
background weak lensing shear using galaxies located using photometric
redshifts --- appears to be the most powerful weak lensing
approach \cite{Bernstein04}.

Atmospheric turbulence results in poor spatial resolution for
high-redshift galaxies observed from the ground. Furthermore,
ground-based telescope and camera optics are far from being
diffraction-limited. So great care must be taken in removing the
effects of seeing, tracking errors, and optical aberrations. Stars
provide the reference for removing these effects, but the number of
usable stars is limited, due to both the limited size of the isoplanatic
patch within which seeing is correlated and difficulty in
distinguishing the few stars from the large number of galaxies at faint
magnitudes. Errors in these corrections set a limit on the quality that
ground-based weak lensing will be able to achieve.

In comparison, space-based wide-field observations resolve many more
sources and the remaining corrections for tracking and telescope
aberrations are much smaller relative to the sizes of the sources.
Figure~4 illustrates the dramatic gain in the number of resolved
galaxies in going from ground- to space-based observations.  Moreover,
in contrast to \hstns, \snap will be placed in a stable thermal
environment so that any corrections can be determined using many
spatially uncorrelated images. Moreover, the deep optical and infrared
photometry from \snap will provide high-quality photometric redshifts
as needed to implement the cross-correlation cosmography method.
Therefore, \snap can provide excellent weak lensing data for
constraining changes in the dark energy equation of state \cite{Rhodes04}.

\begin{figure}
\epsfig{file=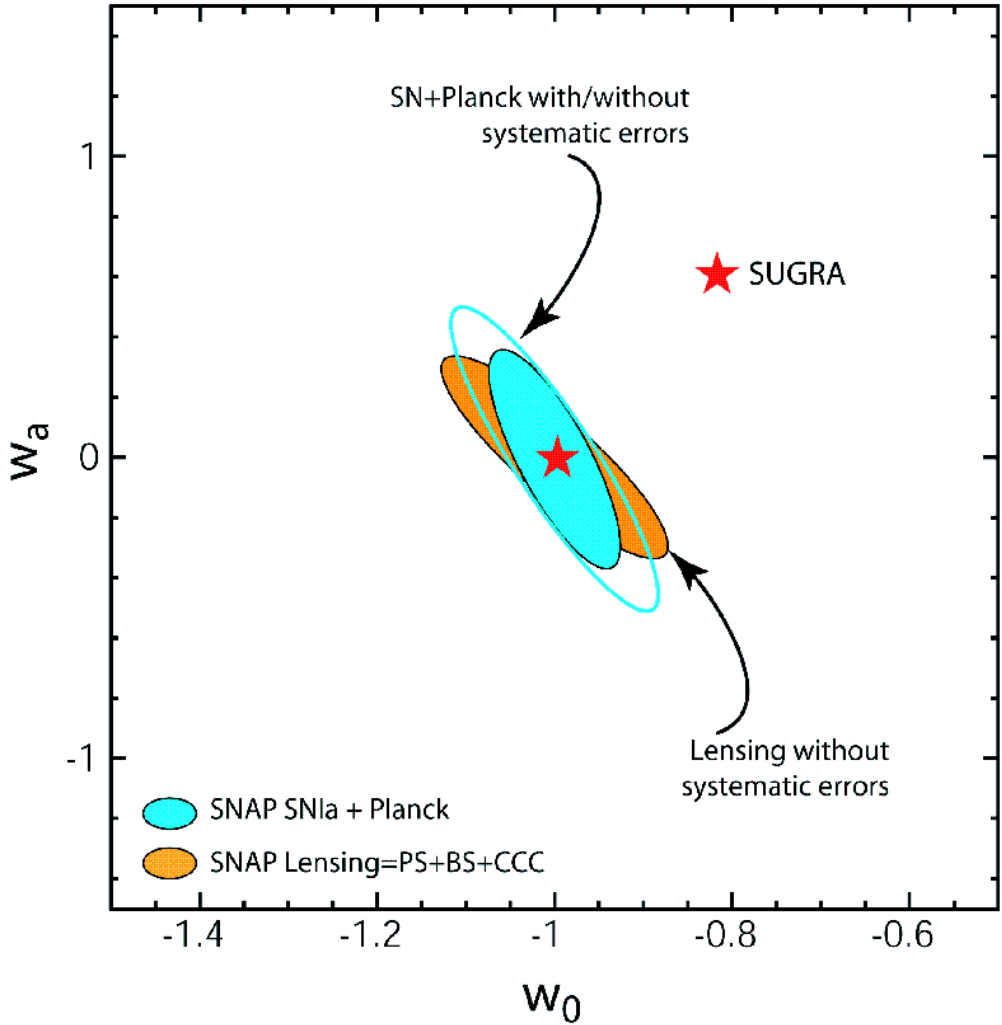,angle=0.0,width=\linewidth}
\caption{Individual constraints on the dark energy equation of state
and its derivative from the \snap Type~Ia SN program and
the weak gravitational lensing program, demonstrating the comparable
and complementary power of these two methods when executed with \snapns.}
\end{figure}

\section{\snap Instrument Concept} 

In detail, the \snap telescope design is a three-mirror anastigmat with
a 2-m primary mirror \cite{Lampton02}. The optics are silver-coated to
maximize the optical and near-infrared throughput and minimize the
thermal emissivity. The imager consists of 0.34 sq.  degrees paved with
red-sensitive CCD's and a scale of 0.10~\arcsec/pixel, and 0.34 sq.
degrees paved with HgCdTe near-infrared detectors with a scale of
0.18~\arcsec/pixel \cite{Lampton02_focalplane}.  The sizes of the
detectors are identical and those made of HgCdTe have a long-wavelength
cutoff which reduces the dark current (and sensitivity to thermal
emission).  These features allow the CCD and HgCdTe detectors to be
mounted in the same focal plane, simplifying the imager construction,
mounting, and alignment. Each detector has its own filter, so a large
filter wheel which would otherwise be needed is avoided. Detailed
simulations show that photometry of sufficient quality and redshift
range is not possible from the ground even under optimistic
assumptions.

\begin{figure}
\epsfig{file=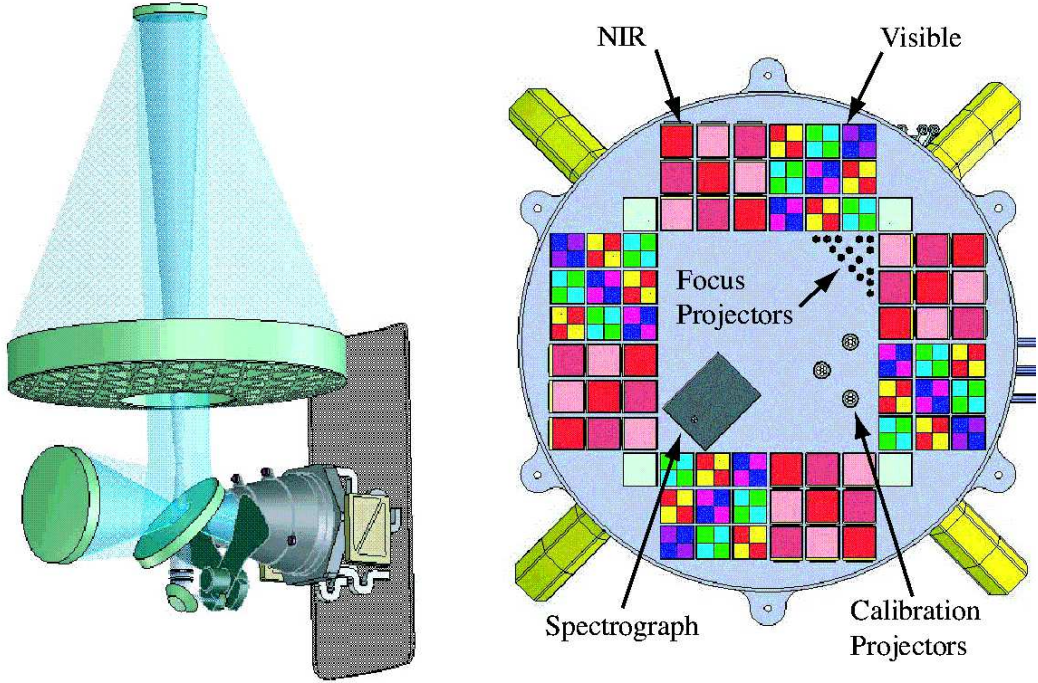,angle=0.0,width=\linewidth}
\caption{Left: Layout of the \snap optical telescope assembly
(OTA) \cite{Lampton02}. The telescope, shutter, camera (shielded by the
cone), electronics, and passive radiator are shown. Right: Layout of
the \snap focal plane \cite{Lampton02_focalplane}.  NIR and visible
detectors populate a single focal plane. Each detector has its own
filter. The filters (and hence detectors) are laid out so that as \snap
passes over a field, each location is observed in all filters.  The
\snap spectrograph is mounted behind the imaging focal plane, and is
fed by an IFU in the imaging focal plane \cite{Prieto02}. Detectors for
guiding populate each quadrant, and projectors for focus and
calibration are also integrated into the focal plane.}
\end{figure}

The \snap spectrograph has a prism as its dispersing element, providing
resolution $R\sim100$ across optical and near-infrared wavelengths
\cite{Prieto02}. Higher resolution is unnecessary due to the velocity
broadening intrinsic to the spectra of Type~Ia supernovae.  Again, the
optical detector is a red-sensitive CCD while the near-infrared
detector is HgCdTe with a long-wavelength cut-off near 1.7~$\mu$m. The
spectrograph is fed by an image slicer integral field unit. The field
is 6\arcsec$\times$3\arcsec, wth the latter dimension being divided into 40
slices.  Detailed simulations show quite dramatically that spectroscopy
of comparable quality is not possible from the ground, even with a
thirty meter telescope with adaptive optics.

\begin{table}
\caption{Nominal parameters of the \snap surveys}
\begin{tabular}{|l|c|c|c|c|}
\hline
Survey Mode&Areal Coverage& Depth   &Resolved Galaxy   & Number of         \\[-0.7em]
           &              &         &Surface Density   & Resolved Galaxies \\[-0.7em]
           &(Sq. Degrees) & (AB Mag)&(\# arcmin$^{-2}$)&                   \\
\hline
Deep SNe   &  15          & 30.3    &   250            &  10$^{7.0}$       \\
Wide       &  300--1000   & 27.7    &   100            &  10$^{8.5}$       \\
Panoramic  & 7000--10000  & 26.7    & 40--50           &  10$^{9.0}$       \\
\hline
\end{tabular}
\end{table}

\section{Conclusion} The nature of dark energy is a fundamental
question, which astronomical observations can address. Type~Ia
supernova and weak lensing are the most powerful techniques yet
developed for studying dark energy.  Reaching the requisite statistical
and systematics accuracy with these techniques requires a widefield
imager in space. \snap represents a very advanced concept for such a
widefield imager, having an instrument suite which is both powerful and
versatile. Table~1 details this power for the Type~I supernova
survey, a nominal wide-field weak lensing survey, and a possible
panoramic weak lensing survey. The planned dark energy program for the
Joint Dark Energy Mission (\jdemns) will produce a treasure trove of
data for archival study.  Guest Observer programs that could further
exploit such a powerful instrument are envisioned in the \jdem concept.
The following conference presentations describe many of the scientific
lines of inquiry which would be greatly advanced with a facility such
as \snapns.

\end{document}